# Spatially Encoded Polaritonic Ultra-Strong Coupling in Gradient Metasurfaces with Epsilon-Near-Zero Modes

Enrico Baù[†,1], Andreas Aigner[†,1], Jonas Biechteler[†,1], Connor Heimig[1], Thomas Weber[1], Thorsten Gölz[1], Stefan A. Maier[2,3], and Andreas Tittl*[,1]

[1]Chair in Hybrid Nanosystems, Nano-Institute Munich, Department of Physics, LMU Munich, Germany

[2]School of Physics and Astronomy, Monash University, Clayton, Victoria 3800, Australia.

[3]Department of Physics, Imperial College London, London SW7 2AZ, United Kingdom.

[†]These authors contributed equally to this work

*Corresponding Author

**Abstract**

We introduce a platform to achieve ultra-strong coupling (USC) between light and matter using widely available materials. USC is a light–matter interaction regime characterized by coupling strengths exceeding 10% of the ground state energy. It gives rise to novel physical phenomena, such as efficient single-photon coupling and quantum gates, with applications in quantum sensing, nonlinear optics, and low-threshold lasing. Although early demonstrations in plasmonic systems have been realized, achieving USC in dielectric platforms, which offer lower losses and high Q-factors, remains challenging due to typically low mode overlap between the photonic field and the material resonance. Here we leverage dielectric dual gradient metasurfaces supporting quasi-bound-states-in-the-continuum to spatially encode both the spectral and coupling parameter space and demonstrate USC to an epsilon-near-zero (ENZ) mode in an ultra-thin $SiO_2$ layer. The strong out-of-plane electric fields in our tapered bar structure overlap exceptionally well with those of the ENZ mode, resulting in a normalized coupling strength of $\eta = 0.101$ and a mode splitting equivalent to 20% of the ENZ mode energy; a four-to-five-fold increase compared to previous approaches. The strong field confinement of our approach opens new possibilities for compact and scalable polaritonic devices, such as tunable frequency converters and low-energy optical modulators.

## Introduction

When the interaction strength between light and matter surpasses a critical threshold, polaritons emerge as quasiparticles that combine the properties of photons with those of material excitations such as excitons or phonons.[1,2] These hybrid states exhibit unique characteristics, including modified dispersion relations,[3,4] strong optical nonlinearities,[5,6] and quantum coherence effects.[7] First realized in microcavities,[8] strong coupling (SC) has recently been observed in various nanophotonic systems,[9,10,11] where high field confinement and small mode volumes were utilized to achieve significant coupling strengths. All-dielectric metasurfaces provide a well-suited platform for enhanced light-matter interaction due to their extraordinary spectral and quality (Q-) factor tunability[12,13], making them highly attractive for various applications, such as nonlinear optics,[14] biosensing,[15] and colorimetry.[16] In particular, the high Q-factors of dielectric metasurfaces supporting symmetry-protected quasi bound-states-in-the-continuum (qBICs) allow clear observation of upper and lower polariton branches demonstrated in recent studies on SC between qBICs and plasmons,[17] excitons,[18,19,20] or molecular vibrations.[21,22]

However, reaching higher coupling strengths would allow a wider spectral range for polaritonic effects, an increased density of photonic states, and potentially stronger nonlinear effects.[23] In particular, the system can reach the ultra-strong coupling (USC) regime when the normalized coupling strength $\eta = \frac{g}{\omega}$ reaches 0.1, where $\omega$ denotes the ground state frequency.[24] Note that the definition of USC does not consider decay rates, but is rather characterized by the coupling strength $g$.[25] While coupling in the USC regime has been successfully demonstrated in optomechanics,[26,27,28] photochemistry,[29] and 2D materials,[30] achieving it in nanophotonic systems is challenging, with initial realizations limited to plasmonic nanogaps.[31,32]

In dielectric metasurfaces, including those typically used for qBICs, $\eta$ has so far remained too low, around five times below the critical threshold for achieving USC,[33] which can be attributed to poor mode overlap between the photonic mode and the resonant material. Unlike plasmonic systems, where strong surface fields drive efficient coupling, the fields in dielectric metasurfaces are typically more delocalized[12]. Using the resonant material as a capping layer, as realized for exciton coupling with 2D materials,[34] or using epsilon-near-zero (ENZ) materials as a substrate layer,[35] has so far been insufficient to reach USC. This issue of low mode overlap is especially apparent in ENZ-based approaches: While ENZ modes are typically confined to thin layers with characteristic out-of-plane field components,[36] the field lines of most photonic modes are predominantly in-plane.[33] As a result, these in-plane fields do not couple to the ENZ mode. Instead, they often couple with other intrinsic loss channels, such as transverse optical (TO) phonons, further reducing the performance of the coupled system.

Here, we present a dual-gradient metasurface platform[37] that achieves USC in the mid-IR by coupling a dielectric qBIC with an ENZ mode. Positioning the highly subwavelength $SiO_2$ layer that supports the ENZ mode at the center of tapered bar resonators enables efficient coupling to electric field vortices of the qBIC. These out-of-plane electric fields lead to a high mode overlap and a splitting of $28.4$ meV for a $114$ nm thick $SiO_2$ layer, corresponding to approximately 20% of the ground state frequency and a normalized coupling strength of $\eta = 0.101$; a four-to-five-fold increase compared to previous approaches with similar ENZ layer thicknesses.[33] The dual-gradient metasurface design allows continuous spatial encoding of both the spectral position and the Q-factor. We experimentally achieve mapping of the anticrossing behavior across 420 individual spectra. By leveraging the continuously tuning of the qBIC resonance from 1400 to 950 $cm^{-1}$, our approach significantly surpasses previous coupling studies in spectral resolution, as earlier works relied on only a limited number of discrete metasurfaces to probe the anticrossing region.[35,38,19] Furthermore, simulations show that our structure already achieves strong coupling with $SiO_2$ films as thin as 5 nm, corresponding to around $\lambda_{ENZ}/1700$.

Additionally, varying the position of the $SiO_2$ layer within the resonator allows to selectively control the coupling strengths of the qBIC mode to either the ENZ and TO phonon mode. With the ENZ layer placed at the center of the tapered bars, the qBIC couples exclusively to it, suppressing interaction with the spectrally close TO phonon. This selectivity contrasts with most existing systems, where the photonic mode couples to both the ENZ mode and the TO phonon,[31] thereby dampening potential hybrid states in between. To our knowledge, such level of control has not been demonstrated in any other phononic system, making our tapered bar gradient particularly well-suited for studying polaritons in highly anisotropic media, such as layered 2D materials with interlayer excitons or phonons in thin films.

## Results

The proposed dual-gradient metasurface consists of continuously tapered Si bars on a $CaF_2$ substrate (**Fig. 1a**). As illustrated in **Fig. 1b**, the unit cell has a pitch $p_x$ along the x-axis and contains two bars of height $h_{res}$ aligned along the y-axis with widths $w_1$ and $w_2$ with $w_1 + w_2 = 0.8\, p_x$. Because the bars continuously span along the whole y-axis of the gradient, the y-pitch $p_y$ can be considered infinitesimal. A thin $SiO_2$ layer of thickness $h_{SiO2}$ serves as the resonant material and is placed at the center of the tapered bars to maximize coupling strength. To break the symmetry and convert the true BIC into a measurable qBIC with a finite Q-factor, an offset between $w_1$ and $w_2$ is introduced. Increasing this offset decreases the radiative Q-factor $Q_{rad}$, following the characteristic relation for symmetry-protected qBICs, $Q_{rad} \propto 1/\alpha^2$, where the asymmetry parameter $\alpha$ is defined as the relative width offset

$$\alpha = \frac{w_1 - w_2}{w_1 + w_2}$$

We select x-polarized light normal to the tapered bars due to the advantageous mode profile of this qBIC. To map both the spectral and coupling space around the ENZ mode, we design dual-gradient metasurfaces (**Fig. 1c**), where the unit cell size continuously increases along the x-axis with a scaling factor $S$ to achieve broad spectral coverage. Additionally, by increasing the asymmetry along the y-axis, the radiative Q-factor can be arbitrarily tuned, limited only by intrinsic losses and fabrication imperfections.

Compared to previous dual gradient or other high Q-factor gradient designs[39,40,37], our unit cell geometry overcomes a key limitation: the misalignment of neighboring unit cells along the scaling direction. In conventional designs, the scaling creates a size mismatch between adjacent unit cells, leading to misalignment between the unit cells, which has been shown to degrade the gradient's performance.[37] Our tapered bar geometry prevents this effect due to the infinitesimal $p_y$, ensuring perfect alignment for neighbouring unit cells (see **Fig. S1**).

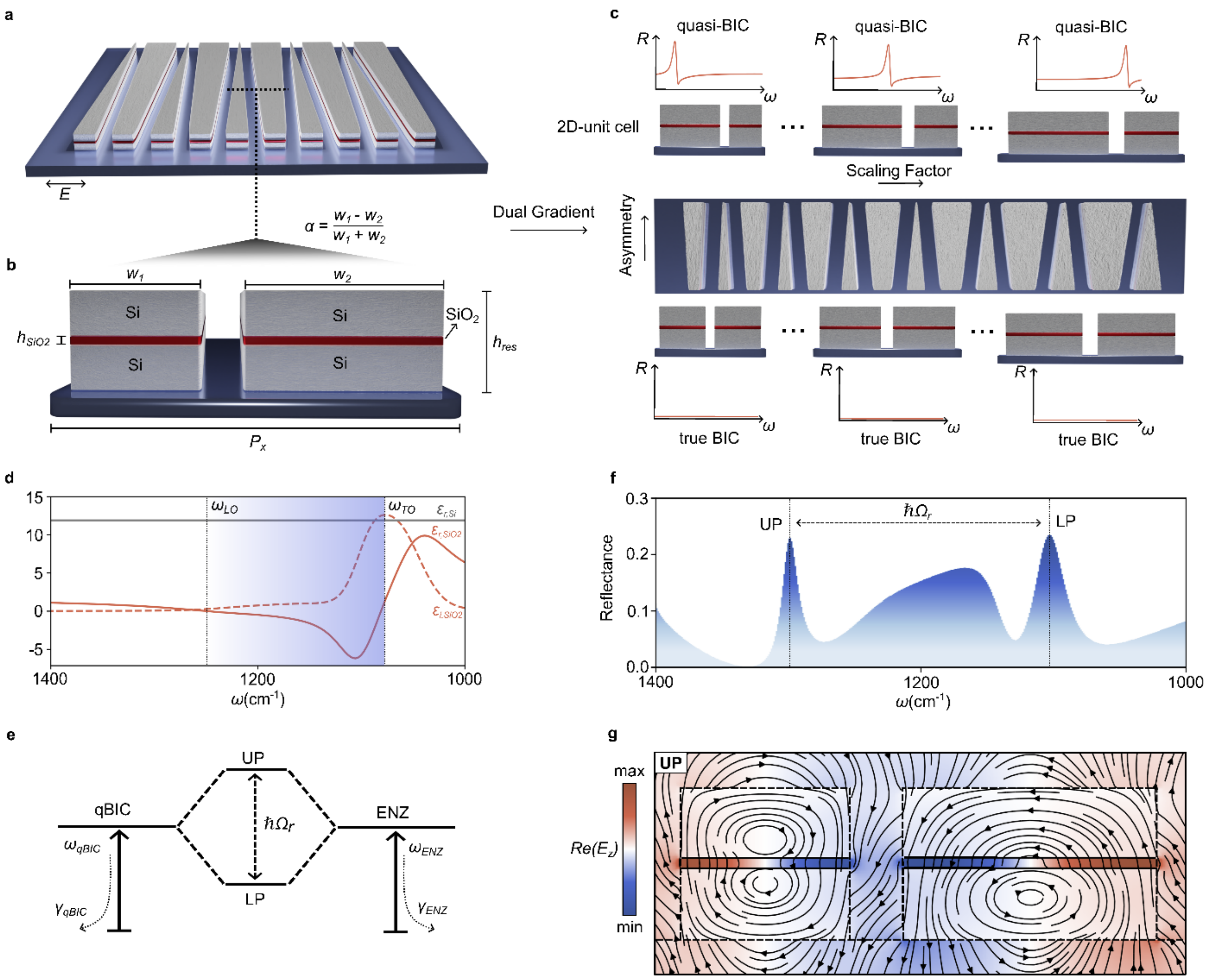

**Figure 1. Tapered bar dual gradient for enhanced vibrational strong coupling. a.** Illustration of a tapered bar metasurface gradient supporting symmetry-protected qBIC resonances. **b.** Schematic of the two-dimensional unit cell composed of Si resonators on a $CaF_2$ substrate, with a thin $SiO_2$ film (thickness $h_{SiO2}$) sandwiched between Si layers. The pitch is given by $p_x$, the height by $h_{res}$, and the combined widths $(w_1 + w_2)$ occupy 80% of $p_x$ (80% filling factor). Their relative offset defines the asymmetry parameter $\alpha$. **c.** Illustration of a dual-gradient design with smoothly varying scaling and asymmetry parameters along both in-plane axes, providing full control over the spectral position and resonance linewidth within a single metasurface. **d.** Real ($\varepsilon_{r,SiO2}$, orange solid curve) and imaginary ($\varepsilon_{i,SiO2}$, orange dashed curve) parts of the adopted $SiO_2$ permittivity.[41] The blue-shaded region denotes the $SiO_2$ reststrahlen band ($\varepsilon_i < 0$). The solid gray curve shows the permittivity of Si, and the TO and LO phonon frequencies are indicated at $\omega_{TO}$ and $\omega_{LO}$, respectively. **e.** Schematic illustrating the strong coupling between a qBIC and an ENZ mode, resulting in a Rabi splitting $\hbar\Omega_r$. **f.** Simulated spectrum showing the splitting of the qBIC mode into UP and LP for $h_{SiO2} = 80$ nm. **g.** Simulated out-of-plane electric field $E_z$ and field lines (black arrows) within a single unit cell at the UP frequency, revealing strong field confinement in the resonant $SiO_2$ layer.

The target wavelength range is set by the dielectric function of $SiO_2$ (**Fig. 1d**), which features a TO phonon and a longitudinal optical (LO) phonon in close proximity to the ENZ region.[41] When the gradient is properly scaled, the qBIC and the ENZ mode couple to form an upper polariton (UP) and a lower polariton (LP), separated by the Rabi splitting $\hbar\Omega_r$, as depicted in the energy level diagram in **Fig. 1e**. An exemplary numerical spectrum for a $SiO_2$ layer with $h_{SiO2} = 80$ nm is shown in **Fig. 1f** and reveals three distinct modes: the UP on the left, the LP on the right, and a broad ENZ mode in between. The asymmetric profile of the ENZ mode can be attributed to the dielectric function of $SiO_2$, which has a small feature around 1200 $cm^{-1}$ due to an asymmetric tension of the out-of-phase Si–O–Si mode.[42] **Fig. 1g** shows the electric field distribution at the UP branch of **Fig. 1f**. The field lines form two vortices of opposite handedness within the tapered bars, revealing the two antiparallel magnetic dipoles characteristic of the qBIC. The color-coded $E_z$ component highlights strong out-of-plane fields within the $SiO_2$ layer, characteristic of ENZ modes. Hence, the observed state clearly combines the features of both the qBIC and ENZ modes, signifying polaritonic behaviour.

Before introducing resonant $SiO_2$ into the gradient, we experimentally analyse pure Si gradient metasurfaces. A Si layer with a height of $1400\ \mathrm{nm}$ was patterned using electron beam lithography to create a $1 \times 3\ \mathrm{mm}^2$ gradient, as depicted in **Fig. 2a**. The scaling factor $S$, and thus the spectral encoding, varies along the x-axis, while the y-axis encodes asymmetries ranging from symmetric at the bottom to highly asymmetric at the top. **Fig. 2b** shows a tilted view scanning electron microscope image for the symmetric and asymmetric cases, respectively.

All measurements were performed using a spectral imaging setup with quantum cascade laser sources (see Methods and **Fig. S2**). Snapshots of the gradient at four distinct wavelengths, captured in reflection mode, are presented in **Fig. 2c**. The resonant areas of the dual gradient (visible as regions of high reflection) show a clear shift toward higher scaling factors along the x-axis as the wavenumber decreases. Additionally, along the y-axis, the high-reflectance region broadens, reflecting the decreasing Q-factor of the qBIC mode with increasing asymmetry factor $\alpha$. The spectra for all $S$ values, measured at three distinct $\alpha$, are shown in **Fig. 2d**. The mode broadens as $\alpha$ increases from 0.07 to 0.21, while the gradient continuously shifts the resonance frequency. Here, $S = 1$ corresponds to a pitch of $p_x = 4000\ \mathrm{nm}$. We note that the wavelength is not shifting perfectly linear with $S$, because the resonator height is constant across the entire gradient and thus the mode volume does not scale linearly with $S$ (**Fig. S3**). Mapping the Q-factor for each individual pixel reveals the Q-factor map in **Fig. 2e**, with spectra taken from the y-axis cut shown in **Fig. 2f**. Their corresponding Q-factors shown in **Fig. 2g** smoothly decrease with $\alpha$ from about 80 to less than 20. The corresponding characteristic antiparallel mode profile is shown in **Fig. S4**.

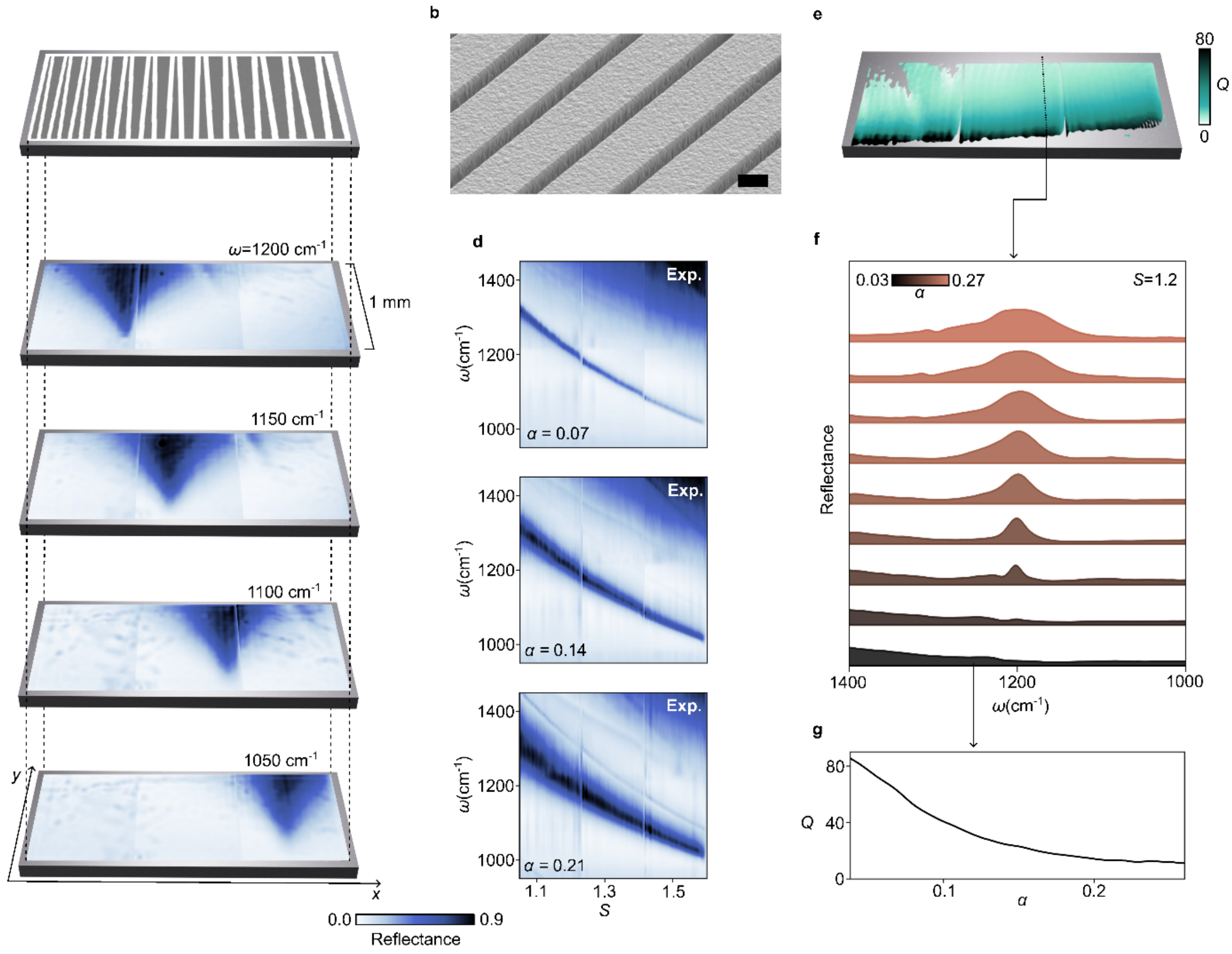

**Figure 2. Optical characterization of the tapered bar dual gradient. a.** Schematic of the dual-gradient design. **b.** SEM image of the fabricated metasurface in tilted view. Scale bar: 1 µm. **c.** Reflectance snapshots of the gradient at four selected wavelengths, where high-reflectance regions indicate a resonance. **d.** Reflectance spectra vs. scaling factor S, extracted along the x-axis of the dual gradient for different asymmetry parameters. **e.** Q-factor map, with each pixel individually fitted. **f.** Reflectance spectra extracted along the y-cut shown in (f). **g.** Corresponding Q-factors.

Next, we insert the $SiO_2$ layer into our design at the center of the Si tapered bars. To probe the coupling between the qBIC and the ENZ mode, we conduct reflectance measurements. The resulting single-wavelength images (**Fig. 3a**) for $h_{SiO2} = 38$ nm, well below the $\lambda/50$ limit in which ENZ modes are excited[36], reveal the coupling when compared sequentially: When approaching the ENZ region, the qBIC disappears due to the energy gap between both polaritonic branches. The qBIC resonance then reappears around $1100\ \mathrm{cm}^{-1}$. This splitting is also evident in the individual spectra plotted in **Fig. 3b**. To determine the coupling parameters of this anticrossing behaviour, we perform measurements of three dual gradients with different $SiO_2$ thicknesses. The spectrally resolved scaling sweeps for $h_{SiO2} = 38$ nm, 76 nm, and 114 nm are shown in **Figs. 3c**, **3d**, and **3e**, respectively. The resulting hybrid modes are fitted (dashed lines) using the well-known coupled-oscillator model,[43] for which the Hamiltonian $H_2$ is given by

$$H_2 = \begin{pmatrix} \omega_{\mathrm{qBIC}} - i\gamma_{\mathrm{qBIC}} & g \\ g & \omega_{\mathrm{ENZ}} - i\gamma_{\mathrm{ENZ}} \end{pmatrix},$$

where $\omega_{\mathrm{qBIC}}$ and $\omega_{\mathrm{ENZ}}$ are the spectral positions and $\gamma_{\mathrm{qBIC}}$ and $\gamma_{\mathrm{ENZ}}$ are the loss rates of the two resonances, respectively. The loss rates used for the following calculations were fitted from the experimental data with a Fano lineshape fit (see **Fig. S5**).

By fitting the eigenvalues of $H_2$ to the UP and LP branches (see **Supplementary Note 1**), we extract the coupling strengths $g$ and the Rabi splitting $\hbar\Omega_{\mathrm{r}}$. Note that while for most SC processes, the spectral position of the polaritonic resonance remains fixed, here, the ENZ mode may shift when $S$ increases because of geometrical alterations, such as changes in resonator width. To account for the non-linear scaling of the qBIC frequency with $S$, we fit both the spectral positions and the linewidths from simulations using the respective film thicknesses, and then incorporate these results into our model. This approach yields more accurate fits by including the complex dielectric function of $SiO_2$. The experimental results for all three $SiO_2$ layer thicknesses are shown as points **(Table S1**) in **Fig. 3f**, alongside the simulated results (solid line, **Table S2**), which are shown in **Fig. S6**. Both sets agree well and show an increasing coupling with $SiO_2$ thickness. To verify whether the measured gradients exhibit SC, we evaluate two standard criteria, $c_1$ and $c_2$,[44] where $c_1 = \Omega_{\mathrm{r}}/(\gamma_{\mathrm{qBIC}} + \gamma_{\mathrm{ENZ}}) > 1$ and $c_2 = g/\sqrt{(\gamma_{qBIC}^2 + \gamma_{ENZ}^2)/2}/> 1$.

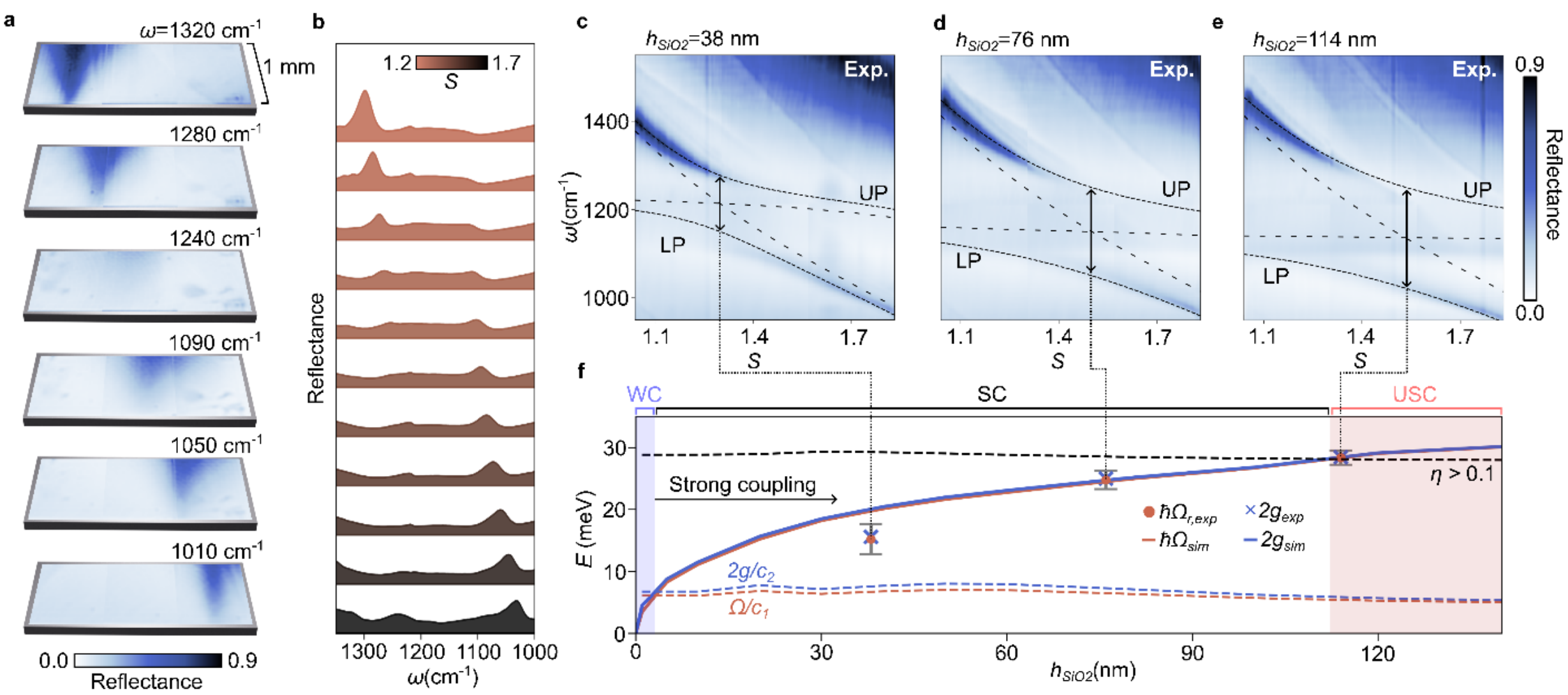

**Figure 3. Experimental ultrastrong coupling between qBIC and ENZ mode. a.** Reflectance images of the metasurface at six different wavelengths, illustrating the strong coupling behavior. **b.** Extracted reflectance spectra for varying scaling factors and an asymmetry parameter of $\alpha = 0.15$, showing splitting of the qBIC mode around the ENZ mode. **c–e.** Reflectance spectra for $SiO_2$ film thicknesses of $h_{SiO2} = 38$ nm, 76 nm, and 114 nm, respectively, showing a clear splitting of the qBIC mode into the upper (UP) and lower (LP) polaritons around the ENZ mode, which increases with film thickness. **f.** Simulated $\hbar\Omega_{\mathrm{r,sim}}$ (solid orange curve) and measured $\hbar\Omega_{\mathrm{r,exp}}$ (orange dots) for the Rabi splitting, as well as the coupling strengths $g_{\mathrm{sim}}$ (solid blue curve) and $g_{\mathrm{exp}}$ (blue crosses) , plotted versus film thickness to highlight enhanced coupling for thicker films. The dashed orange and blue curves represent the strong coupling conditions $c_1$ and $c_2$, while the dashed black line marks the USC regime where $\eta > 0.1$. The blue and red regions delineate the WC and USC regimes respectively.

Both criteria, plotted as dashed lines in **Fig. 3f**, are satisfied by all measurements. Notably, for the thickest measured film ($t = 114$ nm), $c_1 = 5.4$ and $c_2 = 5.0$ correspond to a remarkably large Rabi splitting of $\hbar\Omega_{\mathrm{r}} = 28.4$ meV. This is equivalent to around 20% of the ENZ mode energy, yielding $\eta = 0.101$, which lies within the USC regime. In simulations, the USC regime is clearly reached for film thicknesses of $h_{SiO2} = 120\,\mathrm{nm}$ and $h_{SiO2} = 140\,\mathrm{nm}$. Remarkably, in simulations the transition from weak to strong coupling already occurs for $h_{SiO2} = 5\,\mathrm{nm}$, corresponding to approximately $\lambda_{ENZ}/1700$.

In these measurements, no energy splitting or absorption was observed at the spectral position of the TO phonon (located around 1080 $\mathrm{cm}^{-1}$) [41] clearly indicating that the TO phonon does not interact with the qBIC mode when the $SiO_2$ layer is at the center of the tapered bar. This selective coupling is a result of the qBIC mode profile, first discussed in **Fig. 1d**. There, the two antiparallel magnetic dipoles form an electric field vortex within each tapered bar. Crucially, at the resonator center, in-plane electric fields are nearly absent. Since the atomic displacement for TO phonons is parallel to $E_\mathrm{z}$, it cannot couple to the vortex, resulting in a complete suppression of qBIC–TO coupling (**Fig. 4a**). In contrast, the LO phonons can couple effectively to the qBIC because their atomic displacement is perpendicular to $E_\mathrm{z}$. The electric field vortex within the tapered bars offers unique possibilities for controlling the coupling between the qBIC and the phonon modes. When the $SiO_2$ film is positioned off-center (**Fig. 4b**), both in-plane and out-of-plane electric fields are present, enabling coupling to both the ENZ and TO phonon modes. By following the circular field lines within the tapered bars, it becomes clear that shifting the film away from the center enhances coupling to the TO phonon while reducing interaction with the ENZ mode. When the film is positioned at the very top of the bars (**Fig. 4c**), ENZ coupling is strongly suppressed, allowing only the TO phonon to interact with the qBIC. We can describe this situation using a three-oscillator model with a Hamiltonian $H_3$ of the form

$$H_3 = \begin{pmatrix} \omega_{\mathrm{qBIC}} - i\gamma_{\mathrm{qBIC}} & g_1 & g_2 \\ g_1 & \omega_{\mathrm{ENZ}} - i\gamma_{\mathrm{ENZ}} & 0 \\ g_2 & 0 & \omega_{\mathrm{TO}} - i\gamma_{\mathrm{TO}} \end{pmatrix},$$

where $\omega_{\mathrm{qBIC}}$, $\omega_{\mathrm{ENZ}}$, and $\omega_{\mathrm{TO}}$ denote the resonant frequencies and $\gamma_{\mathrm{qBIC}}$, $\gamma_{\mathrm{ENZ}}$, and $\gamma_{\mathrm{TO}}$ are the corresponding loss rates. Here, $g_1$ and $g_2$ are the coupling strengths, and $\hbar\Omega_{\mathrm{r},1}$ and $\hbar\Omega_{\mathrm{r},2}$ are the respective Rabi splittings for the qBIC–ENZ and qBIC–TO phonon modes. Interactions between the ENZ and TO phonon modes are assumed to be negligible (see **Supplementary Note 1**).

**Figs. 4d, 4e** and **4f** present the corresponding experimental data for various $SiO_2$ film positions. For $z_{\mathrm{SiO2}}/h_{\mathrm{res}} = 0.5$ and $z_{\mathrm{SiO2}}/h_{\mathrm{res}} = 1$, we observe a full suppression of the coupling to the TO phonon and ENZ mode respectively. We obtain an energy splitting of $\hbar\Omega_{\mathrm{r},1} = 28.4$ meV ($\eta = 0.101$) for $z_{\mathrm{SiO2}}/h_{\mathrm{res}} = 0.5$ (**Fig. 4d**) and $\hbar\Omega_{\mathrm{r},2} = 7.0$ meV ($\eta = 0.028$) for $z_{\mathrm{SiO2}}/h_{\mathrm{res}} = 1$ (**Fig. 4f**). In the case of $z_{\mathrm{SiO2}}/h_{\mathrm{res}} = 0.85$, the qBIC mode couples to both the ENZ mode and TO phonon, yielding $\hbar\Omega_{\mathrm{r},1} = 14.2$ meV ($\eta = 0.056$) and $\hbar\Omega_{\mathrm{r},2} = 5.9$ meV ($\eta = 0.024$) (**Fig. 4e**).

Notably, while energy splittings on both modes can be observed experimentally, fully resolving all three energy levels with LP, middle polariton (MP), and UP, is challenging due to the relatively high losses in $SiO_2$. These losses could be mitigated by using alternative phonon-polaritonic materials such as SiC or hBN, which exhibit lower dissipation. The simulated energy splittings (**Fig. 4g** and **Table S3**), together with experimental data (**Table S4**), show that $\hbar\Omega_{\mathrm{r},1}$ drops from 28.4 meV to 0 meV when the $SiO_2$ position is shifted toward the top of the silicon layer. In contrast, $\hbar\Omega_{\mathrm{r},2}$ increases, reaching a maximum splitting of 9.0 meV at $z_{\mathrm{SiO2}}/h_{\mathrm{res}} = 0.9$, as confirmed by numerical simulation results in **Fig. S7**.

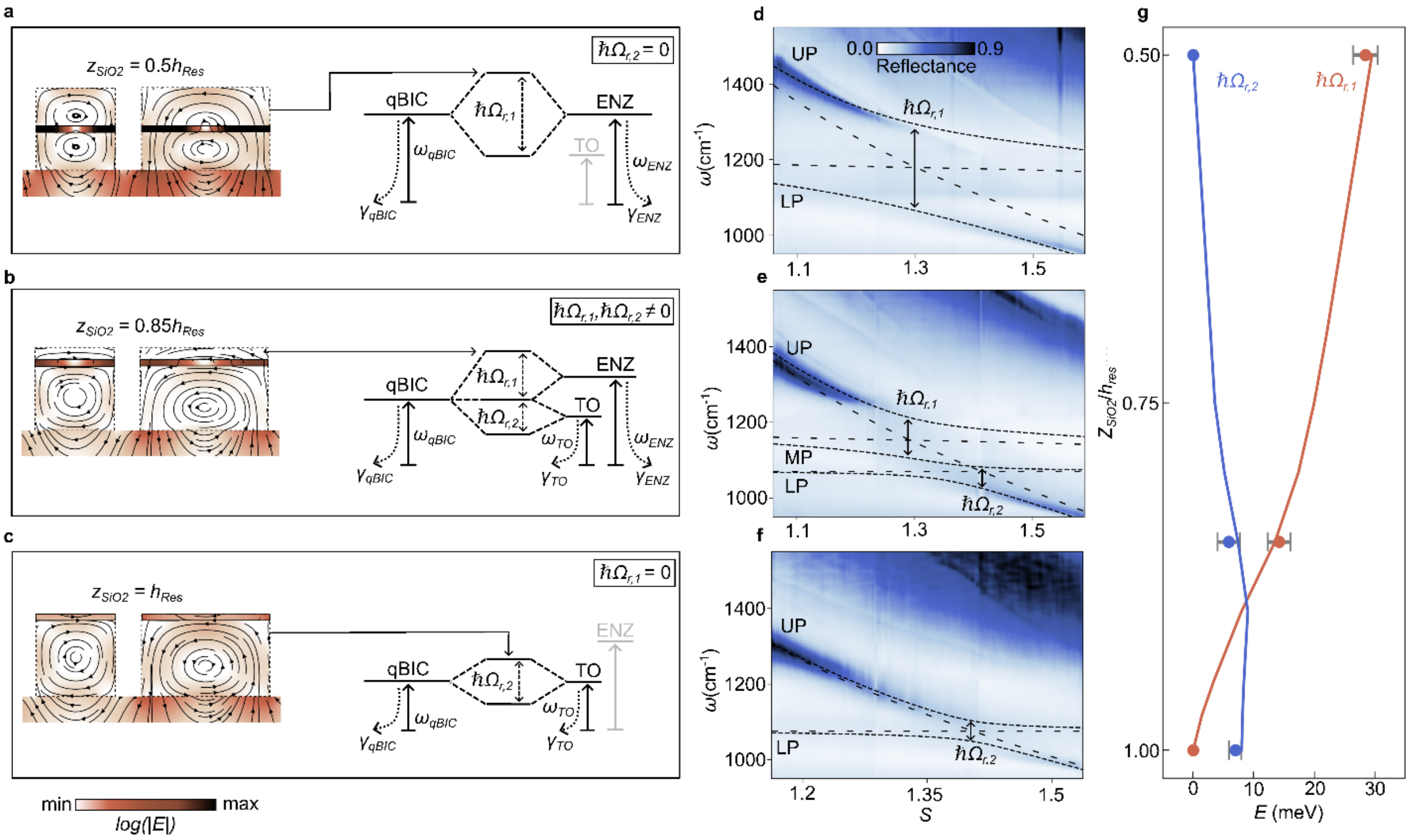

**Figure 4. Experimental realization of selective strong coupling to the TO phonon and ENZ mode. a–c.** Simulated electric field distributions for different $SiO_2$ film positions and corresponding strong coupling diagrams. **a.** When the film is centered, $\hbar\Omega_{r,2} = 0$, indicating no qBIC–TO phonon coupling. **b.** At intermediate positions ($z_{SiO2}/h_{res} = 0.85$), both the ENZ mode and TO phonon couple to the qBIC and $\hbar\Omega_{r,2} \neq 0$. **c.** Moving the film to the top ($z_{SiO2}/h_{res} = 1$) negates the ENZ mode influence and $\hbar\Omega_r = 0$, allowing pure qBIC–TO phonon coupling. **d, e, f.** Experimental demonstration of strong coupling for qBIC–ENZ (d), qBIC–ENZ-TO (e) and qBIC–TO (f). **g.** Simulated (solid curves) and measured (dots) Rabi splittings for qBIC–ENZ (orange) and qBIC–TO (blue), plotted as a function of the $SiO_2$ film position.

Finally, we can maximize the coupling between the $SiO_2$ layer and the qBIC by designing a new gradient structure that efficiently interacts with both the TO phonon and the ENZ mode. This gradient is equivalent to the one shown in **Fig. 3** but includes an additional $SiO_2$ layer at the top, as illustrated in **Fig. S8**. The resulting coupling can be described by both the TO and ENZ modes contributing to an energy splitting of $\hbar\Omega_{r,1,2} = 29.6$ meV (**Fig. S8b**). While the UP and LP are clearly visible, no MP appears. This is likely caused by the dominant qBIC–ENZ coupling, which hybridizes the resonance beyond the spectral position of the TO phonon. Nevertheless, the observed energy splitting remains significantly larger than the pure qBIC–ENZ mode, indicating the influence of the TO phonon. The coupling strength $g$ achieved with this design exceeds 20% of the uncoupled energies of the system, pushing even further into the USC regime.

## Discussion

We successfully demonstrate ultrastrong coupling between an ENZ mode and a qBIC resonance in film thicknesses below the $\lambda/50$ limit in which ENZ modes can still be excited.[36] This extraordinary coupling is made possible by the strong mode overlap between the qBIC and the ENZ mode, outperforming previous designs, as compared in **Fig. S9**. Simulations even indicate that a thickness of 5 nm is sufficient to enter the SC regime, much smaller than in previous designs and deeply subwavelength ($h_{SiO2} < \lambda_{ENZ}/1700$). Our dual-gradient metasurface continuously encodes the Q-factor and spectral position, enabling SC measurements with unprecedented experimental precision. Furthermore, integrating this tapered bar design into gradients resolves misalignment issues between neighboring unit cells, which previously limited the performance of comparable gradient approaches.[39,40,37] Moreover, these gradients are readily transferable to other spectral regions, including the near-IR, where nonlinear effects in materials such as indium tin oxide (ITO) can be utilized.

Furthermore, the combination of extremely small mode volumes and low-loss dielectrics sets our qBIC concept apart from both conventional dielectric and plasmonic implementations of nonlinear ENZ systems: Dielectric designs often exhibit weaker field confinement, while plasmonic systems experience higher losses that diminish the fraction of energy available for nonlinear processes. Such an increase in experimentally achieved coupling strengths paves the way for further exploration of the SC regime and potentially even the deep strong coupling regime.[45,46] In addition, the large splitting and small mode volume offer ideal conditions for active optical tuning of electrically gated ITO,[47,48,49] potentially enabling broad spectral tunability and control over the SC regime.

Lastly, we demonstrate selective coupling of qBICs to either the TO or ENZ mode, a unique feature of our design. We can experimentally suppress either the TO phonon or the ENZ contribution simply by adjusting the position of the film within the resonator. This strategy is valuable not only for phononic materials but also for transition metal dichalcogenides (TMDCs) to isolate specific excitonic resonances or interband transitions. Crucially, the distinctive and by height tunable field distribution in the resonator could facilitate the generation of interlayer excitons[50] in stacked TMDC metasurfaces, which require strong out-of-plane electric field enhancements for efficient excitation. Finally, the excellent mode overlap and high field confinement achieved in our platform offers a direct pathway towards the realization of tunable frequency converters and optical modulators, which may require strong nonlinear effects to function efficiently.

## Materials and Methods

*Numerical Methods*

We conducted the simulations using CST Studio Suite (Simulia), a commercial finite element solver. The setup included adaptive mesh refinement and periodic boundary conditions in the frequency domain. $SiO_2$ was modeled according to the data provided by Franta et al.,[41] and the refractive index of Si and $CaF_2$ were set to 3.32 and 1.43, respectively.

*Fabrication*

For the coupled qBIC-ENZ gradients, fabrication began with the sequential deposition of amorphous silicon via plasma-enhanced chemical vapor deposition (PECVD) using a PlasmaPro 100 system (Oxford Instruments). This was followed by radio frequency sputtering of $SiO_2$ using an Amod PVD system (Angstrom Engineering) and a second silicon deposition step. Finally, a chromium layer was sputtered using the Amod PVD system. The films were spin-coated with a positive electron beam resist, CSAR 62 (Allresist), and patterned via electron beam lithography using an eLINE Plus system (Raith) at 20 kV with a 30 μm aperture. The developed patterns were processed in an amyl acetate bath, followed by a MIBK:IPA (1:9 ratio) bath. Reactive ion etching (RIE) was performed using the PlasmaPro 100 system (Oxford Instruments) to selectively etch the film layers in the following order: chromium (serving as a hard mask after the resist layer was removed), silicon, $SiO_2$, silicon, and finally chromium to strip the remaining hard mask. The pure silicon gradient, the gradient with $SiO_2$ on top, and the version with silicon both on top and at the center were fabricated following the same procedure, with adjustments only to the deposition and etching sequence. The fabrication workflow is visualized in **Fig. S10**.

*Optical characterization*

Optical measurements were conducted using a Spero mid-infrared spectral imaging microscope (Daylight Solutions Inc.), as shown in **Fig. S2**. The system included a 4× magnification objective (NA = 0.15), providing a 2 mm² field of view with a resolution of 480 × 480 pixels and a pixel size of approximately 4 × 4 µm². It was equipped with three tunable quantum cascade lasers, covering wavelengths from 5.6 to 10.5 µm with a spectral resolution of 2 $cm^{-1}$. The lasers emitted linearly polarized light, which was crucial for our measurements. To capture the full 1 × 3 mm² gradient, three spectral images were acquired and stitched together during post-processing.

## Acknowledgments

This project was funded by the Deutsche Forschungsgemeinschaft (DFG, German Research Foundation) under grant numbers EXC 2089/1–390776260 (Germany's Excellence Strategy) and TI 1063/1 (Emmy Noether Program), the Bavarian program Solar Energies Go Hybrid (SolTech), and Enabling Quantum Communication and Imaging Applications (EQAP), and the Center for NanoScience (CeNS). Funded by the European Union (ERC, METANEXT, 101078018 and EIC, NEHO, 101046329). Views and opinions expressed are however those of the author(s) only and do not necessarily reflect those of the European Union or the European Research Council Executive Agency. Neither the European Union nor the granting authority can be held responsible for them. S.A.M. additionally acknowledges the Lee-Lucas Chair in Physics.

## Author contributions

E.B., A.A. and A.T. conceived the idea and planned the research. A.A., J.B. and C.H. contributed to the sample fabrication. E.B., A.A., J.B. and T.G. performed the measurements. E.B. conducted the numerical simulations. E.B., A.A., J.B. and T.W. contributed to the data processing. E.B., A.A., J.B., T.W., S.A.M. and A.T. contributed to the data analysis. S.A.M. and A.T. supervised the project. All authors contributed to the writing of the paper.

## Competing interests

The authors declare that they have no competing interests.

**Data and materials availability**

All data are available in the main text or the supplementary materials.

# Supporting Information

**Spatially Encoded Polaritonic Ultra-Strong Coupling in Gradient Metasurfaces with Epsilon-Near-Zero-Modes**

*Enrico Baù*†*, Andreas Aigner*†*, Jonas Biechteler*†*, Connor Heimig, Thomas Weber, Thorsten Gölz, Stefan A. Maier, and Andreas Tittl**

†These authors contributed equally to this work

*Corresponding Author

**Supplementary Note 1: Coupled oscillator model**

To determine the coupling strengths from simulated and experimental data, a coupled oscillator model was utilized. Since the contribution of the TO phonon can be ignored when the $SiO_2$ layer is sandwiched between the resonator material, the system studied in **Fig. 3** can be described by a 2x2 hamiltonian $H_2$ (as shown in the main manuscript). By diagonalizing $H_2$, we can obtain two distinct parabolic branches $\omega_\pm$ which describe the spectral positions of both hybrid states, called upper polariton and lower polariton:

$$\omega_\pm = \frac{\omega_{qBIC} + \omega_{ENZ}}{2} + \frac{i(\gamma_{qBIC} + \gamma_{ENZ})}{2} \pm \sqrt{g^2 - \frac{1}{4}(\gamma_{qBIC} - \gamma_{ENZ} + i(\omega_{qBIC} - \omega_{ENZ}))^2}$$

We fitted this expression to results and simulations shown in **Fig. 3** to retrieve the coupling strength *g*. At zero detuning ($\omega_{qBIC} = \omega_{ENZ}$), the rabi splitting can be obtained by the following expression:

$$\Omega_r = 2\sqrt{g^2 - \frac{1}{4}(\gamma_{qBIC} - \gamma_{ENZ})^2}$$

To study the coupling between TO phonon, ENZ mode and qBIC mode shown in **Fig.4**, we utilized a three harmonic oscillator model, described by a Hamiltonian $H_3$. The procedure to determine the two coupling strengths $g_1$ between qBIC and ENZ mode and $g_2$ between qBIC and TO phonon is equivalent to the two coupled oscillator model shown above, except that three parabolic branches (UP, MP and LP) are fitted to extract all coupling strengths simultaneously.

**Supplementary Tables**

| $h_{SiO2}$(nm) | $\Omega_r$(meV) | $\Delta\Omega_r$(meV) | g(meV) | $\omega_{ENZ}$($cm^{-1}$) | $\gamma_{ENZ}$($cm^{-1}$) | $\eta$ |
|---|---|---|---|---|---|---|
| 38 | 15.2 | ±2.5 | 7.8 | 1182 | 41 | 0.053 |
| 76 | 24.8 | ±1.5 | 12.5 | 1152 | 39 | 0.875 |
| 114 | 28.4 | ±1.2 | 14.2 | 1135 | 31 | 0.101 |

**Table S1: Experimental coupling strengths and energy splittings for various SiO2 layer thicknesses.**

| $h_{SiO2}$(nm) | $\Omega_r$(meV) | $\Delta\Omega_r$(meV) | $g$(meV) | $\omega_{ENZ}$(cm$^{-1}$) | $\gamma_{ENZ}$(cm$^{-1}$) | $\eta$ |
|---|---|---|---|---|---|---|
| 1 | 3.4 | ±0.2 | 2.2 | 1164 | 36 | 0.016 |
| 3 | 5.9 | ±0.2 | 3.3 | 1164 | 36 | 0.023 |
| 5 | 8.3 | ±0.3 | 4.4 | 1164 | 36 | 0.030 |
| 10 | 11.1 | ±0.3 | 5.7 | 1164 | 36 | 0.040 |
| 20 | 15.2 | ±0.5 | 7.8 | 1168 | 42 | 0.054 |
| 30 | 18.2 | ±0.6 | 9.3 | 1184 | 39 | 0.063 |
| 40 | 20.1 | ±0.8 | 10.2 | 1181 | 42 | 0.070 |
| 50 | 21.7 | ±1.1 | 11.0 | 1173 | 44 | 0.076 |
| 60 | 22.8 | ±1.4 | 11.6 | 1164 | 43 | 0.080 |
| 80 | 25.0 | ±1.6 | 12.6 | 1149 | 39 | 0.088 |
| 100 | 26.7 | ±1.8 | 13.4 | 1139 | 33 | 0.095 |
| 120 | 29.1 | ±1.9 | 14.6 | 1133 | 30 | 0.104 |
| 140 | 30.1 | ±2.0 | 15.1 | 1129 | 28 | 0.108 |

**Table S2: Simulated coupling strengths, energy splittings and decay rates for various SiO2 layer thicknesses.**

| $z_{SiO2}/h_{res}$ | $\Omega_{r,1}$(meV) | $\Omega_{r,2}$(meV) | $g_1$(meV) | $g_2$(meV) |
|---|---|---|---|---|
| 0.5 | 29.5 | 0 | 14.8 | 0 |
| 0.7 | 22.1 | 2.9 | 11.4 | 2.0 |
| 0.75 | 20.0 | 3.6 | 10.4 | 2.3 |
| 0.8 | 17.4 | 5.2 | 9.2 | 2.9 |
| 0.85 | 13.5 | 7.4 | 7.4 | 3.9 |
| 0.9 | 8.0 | 9.0 | 5.0 | 4.7 |
| 0.95 | 3.4 | 8.4 | 3.5 | 4.4 |
| 0.975 | 1.4 | 8.1 | 3.0 | 4.3 |
| 1 | 0 | 8.0 | 0 | 4.2 |

**Table S3: Simulated coupling strengths and energy splittings for various SiO2 layer positions.**

| $z_{SiO2}/h_{res}$ | $\Omega_{r,1}$(meV) | $\Delta\Omega_{r,1}$(meV) | $\Omega_{r,2}$(meV) | $\Delta\Omega_{r,2}$(meV) | $g_1$(meV) | $g_2$(meV) |
|---|---|---|---|---|---|---|
| 0.5 | 28.4 | ±1.2 | 0 | - | 14.3 | 0 |
| 0.85 | 14.2 | ±1.9 | 5.9 | ±1.9 | 8.2 | 3.2 |
| 1 | 0 | - | 7.0 | ±1.0 | 0 | 3.7 |

**Table S4: Experimental couling strengths and energy splittings for various SiO2 layer positions.**

## Supplementary Figures

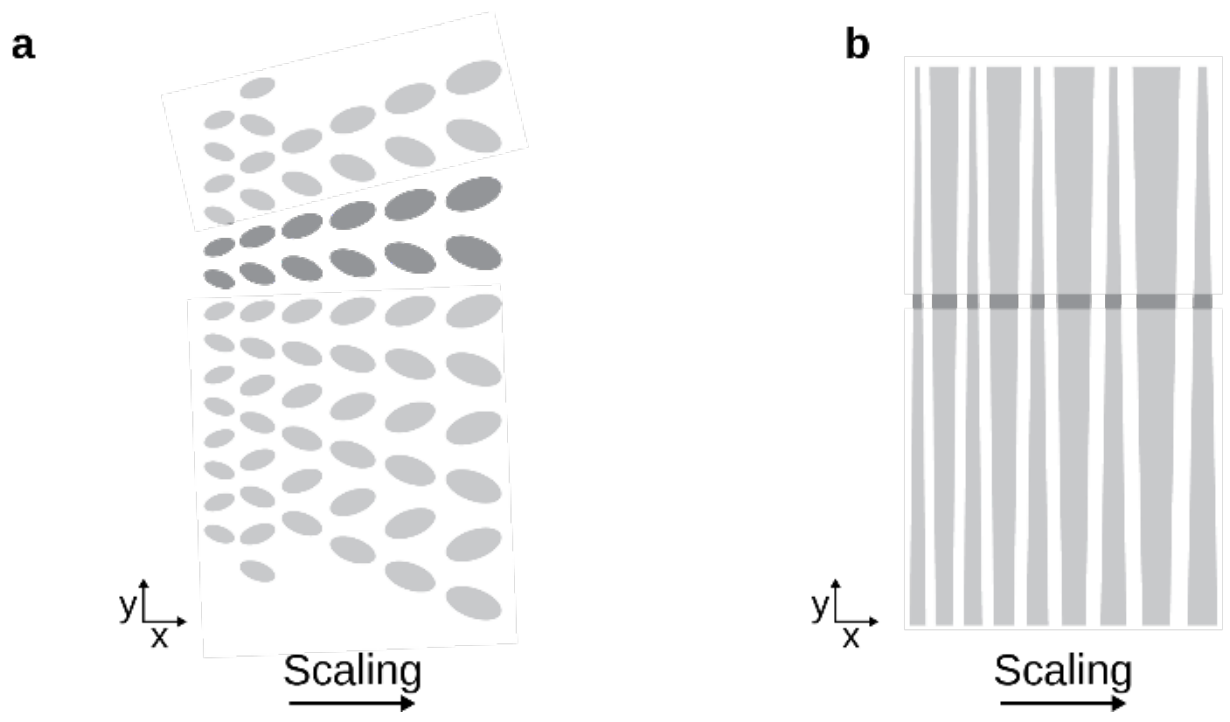

**Figure S1: Unit cell alignment. a.** Conventional high Q-factor scaling gradient design with scaling along the x-axis.[33,34,31] In this design, the y-axis pitch increases between neighbouring columns of resonators, disturbing the alignment within a row. This misalignment is evident in the highlighted row, where the y-center positions shift with scaling. **b.** Tapered-bar gradient with an infinitesimal y-axis pitch and scaling along the x-axis. This configuration enables perfect alignment of neighbouring unit cells along the x-axis, a substantial improvement over previous designs.

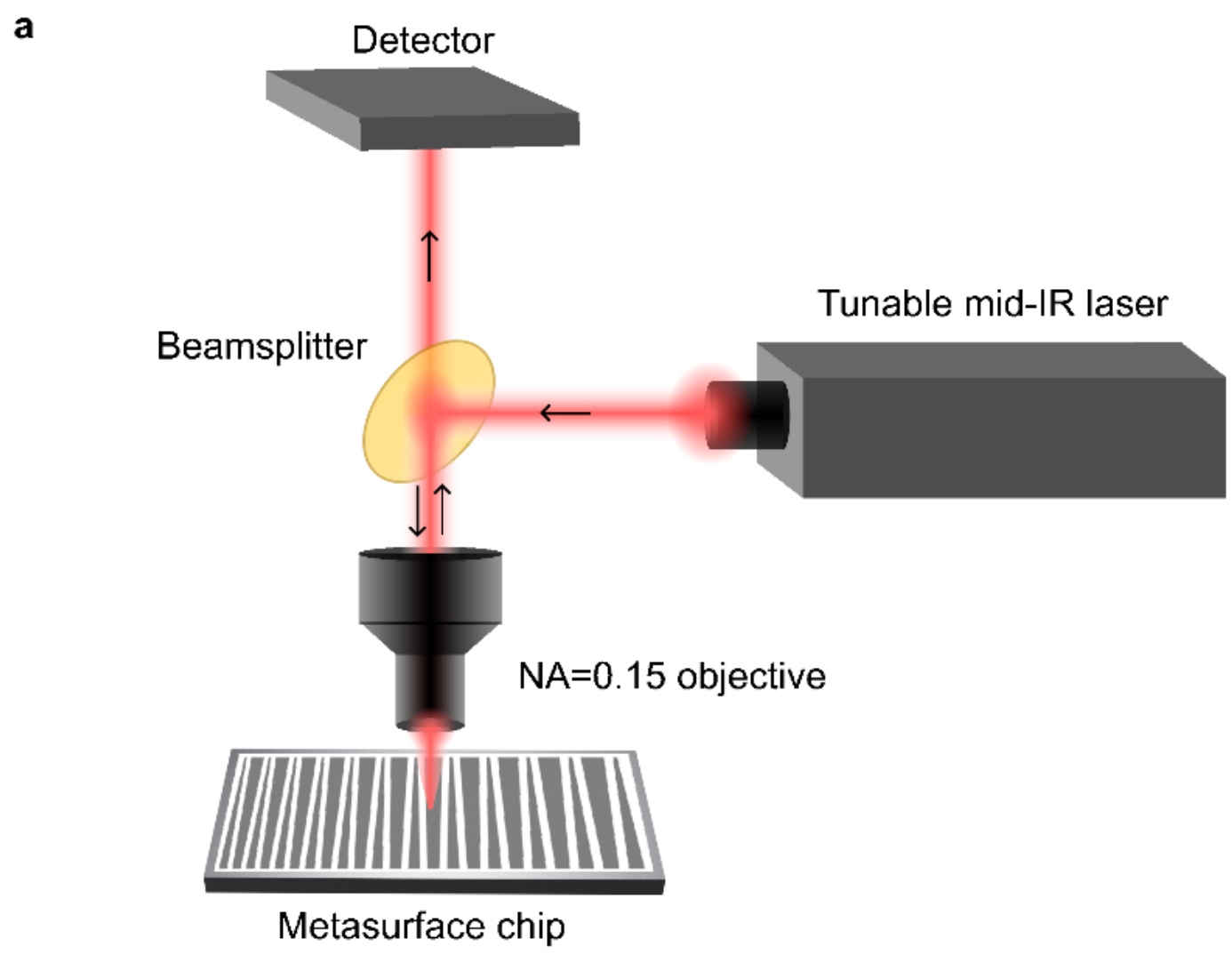

**Figure S2: Sketch of the measurement setup. a.** A tunable mid-IR laser is guided via a dichroic mirror onto the sample through a high N.A. ($N.A.$ = 0.15) objective. The reflected light is collected by the same objective and projected onto a 480x480 pixel detector. The lasers can be tuned in 2 cm$^{-1}$ steps across the target spectral range. This results in an image being captured at each wavelength, allowing for hyperspectral imaging.

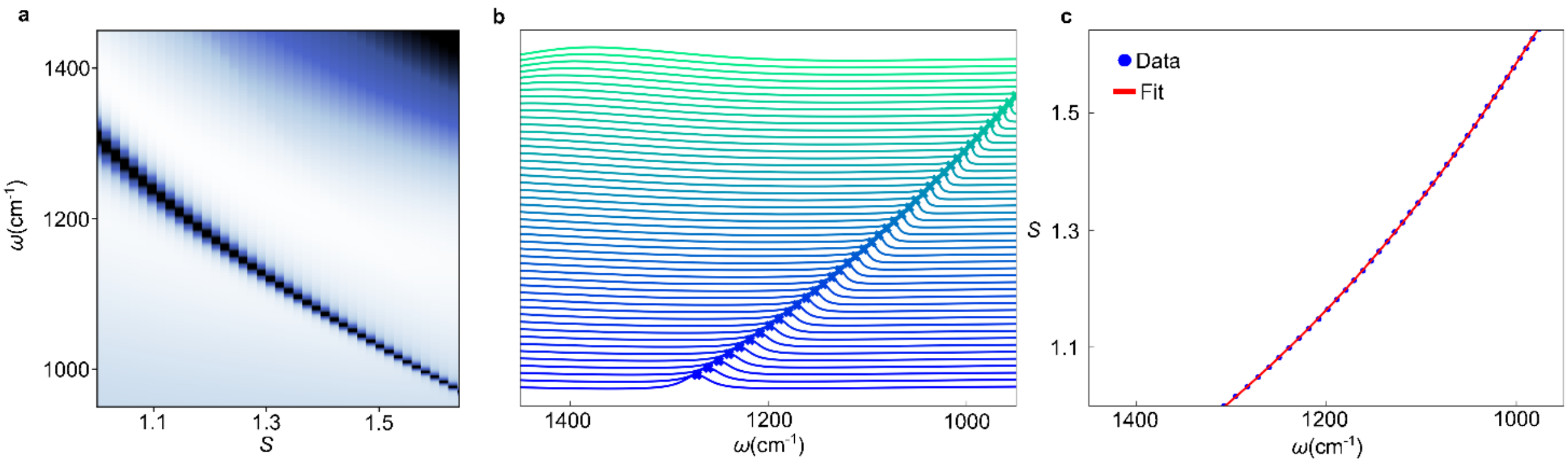

**Figure S3: Simulations of Si metasurface without $SiO_2$. a.** Simulated reflectance spectra vs. scaling factor. **b**. Individual spectra across the target wavelength range. **c.** Fitted peaks vs. scaling factor, fitted with a cubic function of the form $ax^3 + bx^2 + cx + d$ to capture the nonlinear behavior of the qBIC mode resonance position.

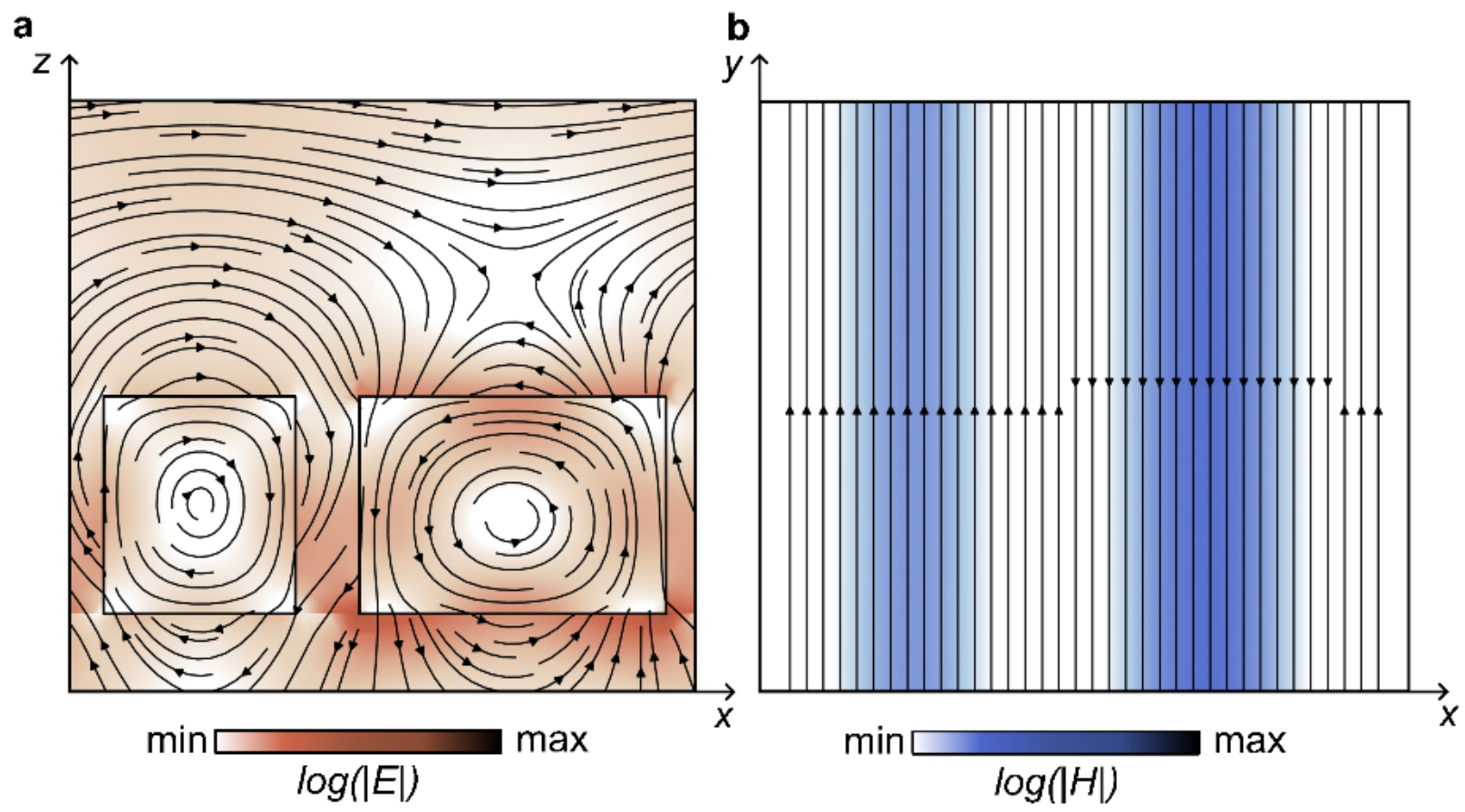

**Figure S4: Simulated *E*- and *H*-fields of pure Si tapered bar metasurface. a.** Cross-section of a single unit cell showing electric field lines forming a vortex inside the resonator at the qBIC resonance.. **b.** Top view of a single unit cell showing the magnetic dipole that is generated at the qBIC resonance.

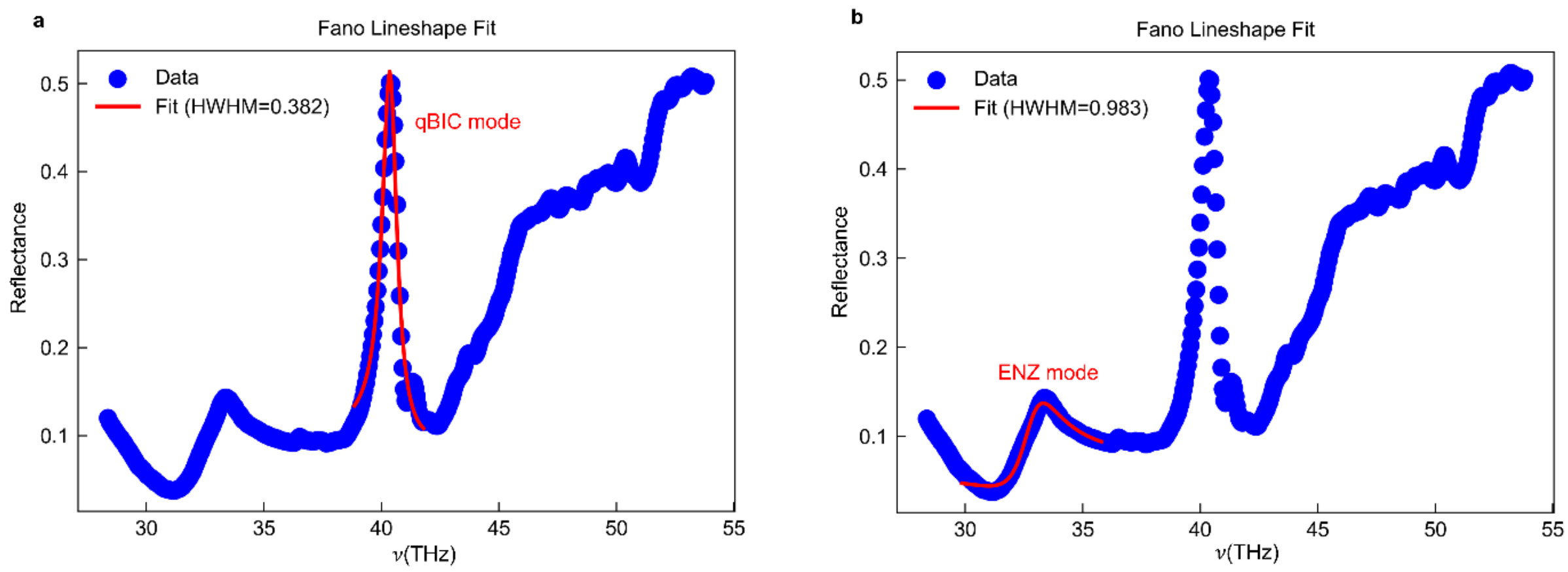

**Figure S5: Resonance linewidth fitting a, b.** Experimental fano lineshape fitting of qBIC and ENZ mode respectively ($h_{SiO2}$ = 114 nm). The fits yield a HWHM of 0.382 THz for the qBIC mode and 0.983 THz for the ENZ mode.

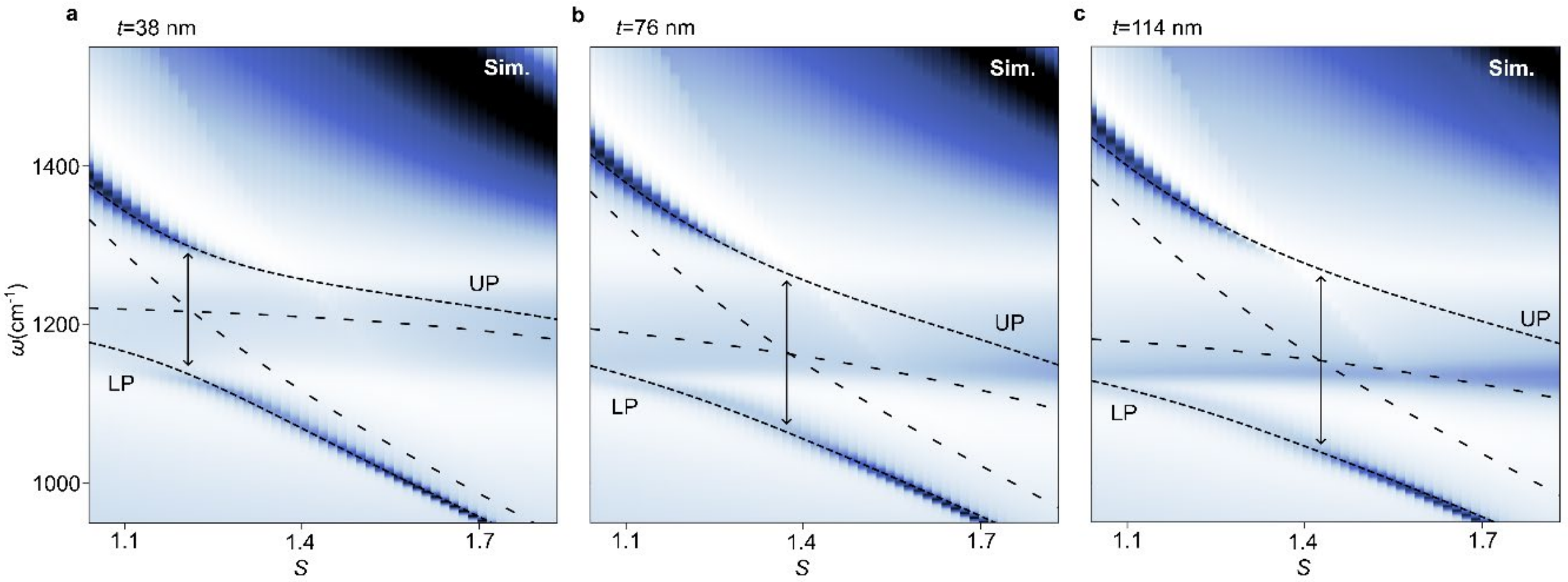

**Figure S6: Simulations of Si metasurface with $SiO_2$ layer of varying thicknesses. a, b, c.** Simulated reflectance spectra vs. scaling factor for $SiO_2$ thicknesses of $h_{SiO2}$=38, 76 and 114 nm respectively. The strong coupling fits are shown as dashed (UP and LP) and loosely dashed (qBIC, ENZ mode positions) curves.

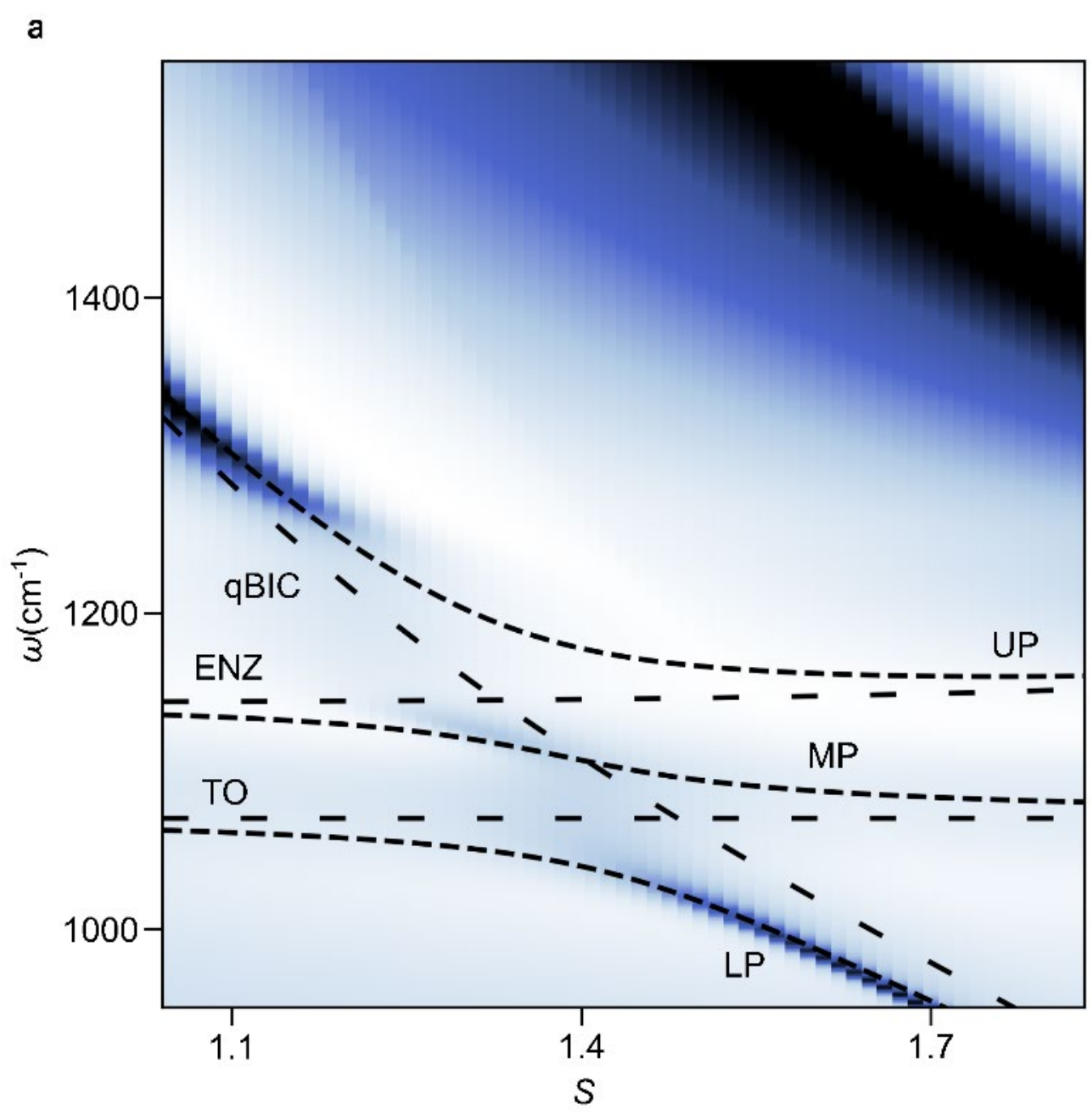

**Figure S7: Simulations of Si metasurface with $SiO_2$ layer at position $z_{SiO2}$=0.9$h_{res}$. a.** Simulated reflectance spectra vs. scaling factor for ENZ position at $z_{SiO2}$=0.9$h_{res}$. The strong coupling fits obtained from the triple coupled oscillator model are shown as dashed (UP, MP and LP) and loosely dashed (qBIC, EN, TO) curves.

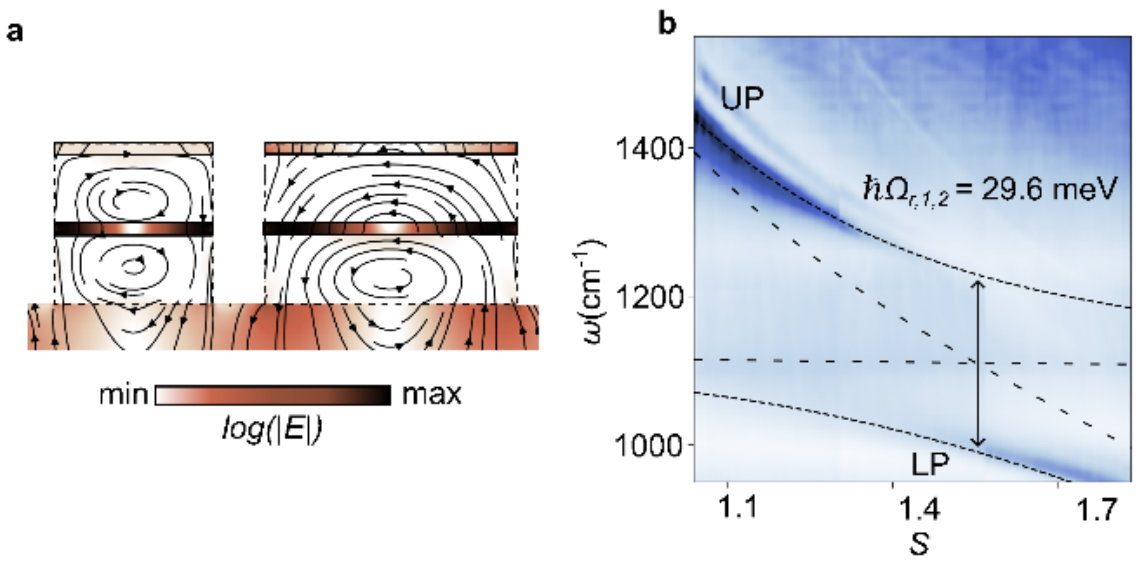

**Figure S8. Ultra-strong coupling of a qBIC mode to an ENZ mode and a TO phonon. a.** Simulated electric field distribution for a metasurface design featuring two $SiO_2$ layers to maximize the combined qBIC–ENZ and qBIC–TO coupling. **b.** Experimental realization of ultra-strong coupling in the system shown in (a), exhibiting a Rabi splitting of 29.6 meV.

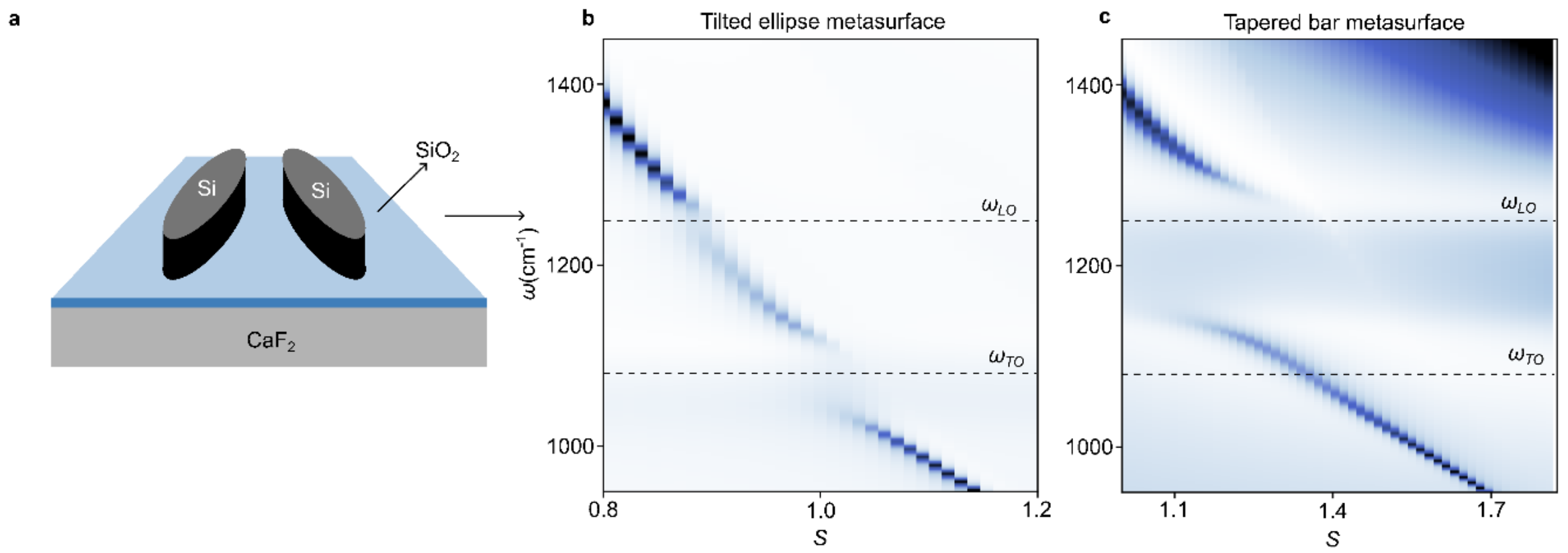

**Figure S9: Comparison of conventional approach and our proposed tapered bar design. a.** Sketch of a tilted ellipse metasurface consisting of Si ellipse resonators, with the resonant $SiO_2$ layer placed on top of the substrate, similar to proposed designs.[30] **b,c.** Simulated reflectance spectra of tilted ellipse metasurface shown in a and our proposed tapered bar metasurface for the same thickness of $SiO_2$ ($h_{SiO2}$ = 80 nm). Compared to the tilted ellipse metasurface, the tapered bar metasurface shows much larger energy splitting around the ENZ mode, while fully supressing coupling between qBIC and TO phonon. $P_x$ = 4000 nm for $S$ = 1.

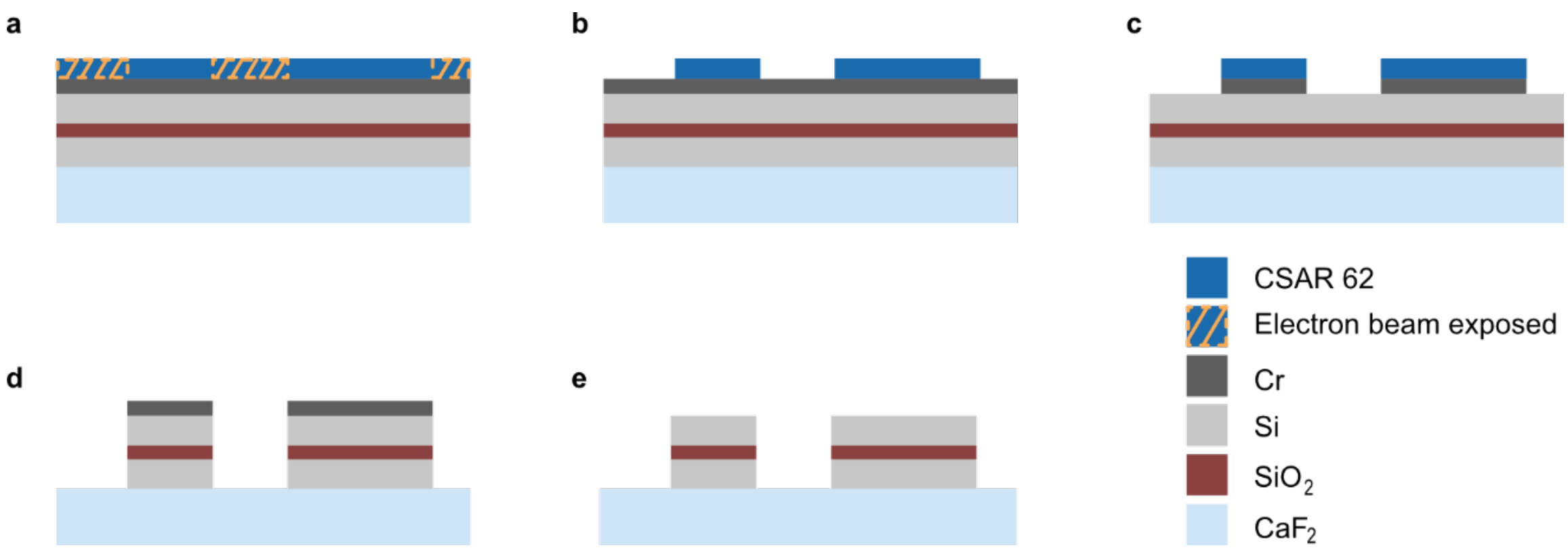

**Figure S10: Fabrication Workflow a.** Stack of Si, $SiO_2$, Si, Cr and CSAR 62 resist on a $CaF_2$ substrate after electron beam exposure in gap areas. **b.** Development. **c.** Cr reactive ion etching. **d.** Reactive ion etching of Si, $SiO_2$ and again Si. **e.** Cr mask removal in final dry etching step.